\documentclass{article}\begin{document}
\centerline{\bf A Lattice Spanning-Tree Entropy Function }

\vskip .5in
\centerline{ M.L. Glasser}
\vskip .2in
\centerline{Departamento de Fisica Teorica, Atomica y Optica, Universidad de Valladolid}
\centerline{47071 Valladolid, Spain} 
\vskip .1in
\centerline{Center for Quantum Device Technology, Clarkson University, Potsdam}
\centerline{ NY 13699-5820, USA}\vskip .1in
\centerline{George Lamb}
\centerline{2942 Ave. del Conquistador}
\centerline{Tucson, AZ 85749-9304, USA}

\vskip .4in
\begin{quote}
\noindent {\bf Abstract} The function
$$W(a,b)=\int_0^{2\pi}dx\int_0^{2\pi}dy\ln[1-a\cos\; x-b\cos\; y-(1-a-b)\cos(x+y)]$$ which expresses the spanning-tree entropy for various two dimensional lattices, for example, is evaluated directly in terms of
standard functions. It is applied to derive several limiting values of the Triangular lattice Green function.\end{quote}

\vskip .6in\noindent
PACS: 02.30.-f, 05.50+q

\newpage
\noindent {\bf Introduction}\vskip .1in

The function
$$W(a,b)=\int_0^{2\pi}dx\int_0^{2\pi}dy\ln[1-a\cos\; x-b\cos\; y-c\cos(x+y)]$$
with $a+b+c=1$
arises frequently in the statistical physics and combinatorics of two dimensional lattice systems. For example:
\begin{quote}
\noindent (i) $a=b=1/2$
$$S_{sq}=\frac{\ln(2)}{2\pi^2}+\frac{W(a,b)}{4\pi^2}$$
is the spanning-tree entropy for the square lattice[1].

\noindent (ii) $a=b=1/3$
$$S_{tr}=\frac{\ln(6)}{4\pi^2}+\frac{W(a,b)}{4\pi^2}$$
is the spanning-tree entropy for the triangular lattice[2].

\noindent
(iii)$$a=\frac{\sinh\; K_1}{\sinh\; K_1+\sinh\; K_2+\sinh\; K_3}$$
$$b=\frac{\sinh\; K_2}{\sinh\; K_1+\sinh\; K_2+\sinh\; K_3}$$
$$F_I=\ln(2)+\frac{1}{8\pi^2}\ln[\sinh\; K_1+\sinh\; K_2+\sinh\; K_3]+\frac{W(a,b)}{8\pi^2}$$
is the critical free energy of the Ising model on a triangular lattice[3].\end{quote}

By comparing the free energies of various Potts models which are known to be related and have been worked out in different ways, Chen and Wu[4]  have proposed that
$$\frac{1}{4\pi^2}\int_0^{2\pi}d\theta\int_0^{2\pi}d\phi\ln[A+B+C-A\cos\theta-B\cos\phi-C\cos(\theta+\phi)]$$
$$=-\ln(2S)+\frac{2}{\pi}[Ti_2(AS)+Ti_2(BS)+Ti_2(CS)]$$
where $A,B,C\ge0$ and $S=1/\sqrt{AB+BC+CA}$. The aim of this note is to provide a direct proof of this formula. 
\vskip .1in\noindent {\bf Calculation}\vskip .1in

By symmetry, one easily finds
$$W(a,b)=2[W_+(a,b)+W_-(a,b)]\eqno(1)$$
where
$$W_{\pm}(a,b)=\int_0^{\pi}dx\int_0^{\pi}dy\ln[1-a\cos\; x-b\cos\; y-c\cos(x\pm y)].$$
Next, we make the standard change of variable
$$u=tan(x/2)\mbox{\hskip .3in} v=tan(y/2)$$ to obtain
$$W_+(a,b)=4\int_0^{\infty}\int_0^{\infty}\frac{dudv}{(1+u^2)(1+v^2)}\ln[\frac{2}{(1+u^2)(1+v^2)}]$$
$$+4\int_0^{\infty}\int_0^{\infty}\ln[(u+v)^2-a((u+v)^2-u^2(1+v^2))-b((u+v)^2-v^2(1+u^2))]\frac{dudv}{(1+u^2)(1+v^2)}.$$
The first integral is elementary, giving
$$W_+(a,b)=-3\pi^2\ln(2)$$
$$+4\int_0^{\infty}\int_0^{\infty}\frac{\ln[c(u+v)^2+au^2(1+v^2)+bv^2(1+u^2)]}{(1+u^2)(1+v^2)}dudv.$$
Similarly,
$$W_-(a,b)=-3\pi^2\ln(2)$$
$$+4\int_0^{\infty}\int_0^{\infty}\frac{\ln[c(u-v)^2+au^2(1+v^2)+bv^2(1+u^2)]}{(1+u^2)(1+v^2)}dudv.$$
 By inserting the last two expressions into (1) and noting that the resulting integrand is even in $v$, we have
 $$W(a,b)=-12\pi^2\ln(2)+8F(a,b),$$
 where
 $$F(a,b)=\int_0^{\infty}\frac{du}{1+u^2}\int_{-\infty}^{\infty}\frac{dv}{1+v^2}\ln[(1-b)u^2+(1-a)v^2+(a+b)u^2v^2+2cuv].$$
 Next, let $u=vw$ to obtain
 $$F(a,b)=2\int_0^{\infty}\frac{udu}{1+u^2}\ln(u)\int_{-\infty}^{\infty}\frac{dw}{1+u^2w^2}$$
 $$+\int_0^{\infty}\frac{du}{u(1+u^2)}\int_{-\infty}^{\infty}\frac{dw}{w^2+u^{-2}}\ln[1-b+(1-a)w^2+(a+b)u^2w^2+2cw].$$
 The first integral vanishes and since [5]
 $$\int_0^{\infty}\frac{dw}{w^2+d^2}\ln[\alpha w^2+2\beta w+\gamma]=\frac{\pi}{d}\ln[\alpha d^2+\gamma+2d\sqrt{\alpha \gamma-\beta^2}],$$
 after the substitution $u=1/z$, \newpage
 $$F(a,b)=$$ $$\pi\int_0^{\infty}\frac{dz}{1+z^2}\ln[(1-a)z^2+(1+a)+2\sqrt{[ac+b(1-b)]z^2+(1-b)(a+b)}].\eqno(2)$$
 
 Now, let us define
 $$A=\frac{y\cot(\theta/2)}{1+\sqrt{1+y^2}}\mbox{\hskip .3in} B=\frac{y\tan(\theta/2)}{1+\sqrt{1+y^2}}.$$
 Then $(A+B)/(1-AB)=y\csc\theta$, so $tan^{-1}(y\csc\theta)=tan^{-1}A+\tan^{-1}B$. However,
 $$\csc\theta\; \tan^{-1}(y\csc\theta)=-\frac{d}{d\theta}\int_1^{\csc\theta}\frac{tan^{-1}(yu)}{\sqrt{u^2-1}}du$$
 $$\csc\theta\; tan^{-1}A=-\frac{d}{d\theta}\int_0^A\frac{tan^{-1}u}{u}du$$
 $$\csc\theta\; tan^{-1}B=\frac{d}{d\theta}\int_0^B\frac{tan^{-1}u}{u}du.$$
 Hence, since both sides vanish for $\theta=\pi/2$,
 $$\int_1^{\csc\theta}\frac{\tan^{-1}(yu)}{\sqrt{u^2-1}}du=Ti_2(A)-Ti_2(B)\eqno(3)$$
 where 
 $$Ti_2(z)=\int_0^z\frac{tan^{-1}x}{x}dx.$$
 With
 $$\theta=\csc^{-1}\sqrt{\frac{b^2-a^2+1}{1-a^2}},\mbox{\hskip .2in} y=\sqrt{a^{-2}-1},\mbox{\hskip .2in}
 u=\sqrt{\frac{x^2-a^2+1}{1-a^2}}$$
 in (3) one obtains
 $$\int_0^b\frac{dx}{\sqrt{x^2-a^2+1}}tan^{-1}\frac{\sqrt{x^2-a^2+1}}{a}=$$ $$
 Ti_2\left(\frac{\sqrt{b^2+1-a^2}+b}{1+a}\right)-Ti_2\left(\frac{\sqrt{b^2+1-a^2}-b}{1+a}\right).\eqno(4)$$
 
 We next consider the integral
 $$g(a,b)=\int_0^{\infty}\frac{ds}{s^2+1}\ln[\sqrt{b^2(s^2+1)+1}+a]$$
 for which it is elementary to determine
 $$g(a,0)=\frac{\pi}{2}\ln(1+a)$$
 $$\frac{\partial}{\partial u}g(a,u)=\frac{\tan^{-1}(\sqrt{u^2+1-a^2}/a)}{\sqrt{u^2+1-a^2}}.$$
 Therefore, by integrating over $u$ using (4), we find
 $$g(a,b)=\frac{\pi}{2}\ln(1+a)+Ti_2\left(\frac{\sqrt{b^2+1-a^2}+b}{1+a}\right)-Ti_2\left(\frac{\sqrt{b^2+1-a^2}-b}{1+a}\right)$$
 which is easily transformed into
 $$\int_0^{\infty}\frac{\ln[\alpha+\sqrt{\beta^2x^2+\gamma^2}]}{x^2+1}dx=\frac{\pi}{2}\ln[\beta+\sqrt{\gamma^2-\alpha^2}]$$
 $$+Ti_2\left(\frac{\alpha+\sqrt{\gamma^2-\beta^2}}{\beta+\sqrt{\gamma^2-\alpha^2}}\right)+Ti_2\left(\frac{\alpha-\sqrt{\gamma^2-\beta^2}}{\beta+\sqrt{\gamma^2-\alpha^2}}\right).\eqno(5)$$
 
 The argument of the logarithm in the integrand of  $F(a,b)$ in (2) can be factored: Let
 $R=\sqrt{[ac+b(1-b)]x^2+(1-b)(a+b)}$; then
 $$(1-a)x^2+1+a+2R=$$
 $$[ac+b(1-b)]^{-1}[a(1-a)+2bc+(1-a)R][a+R].$$
 The integrals resulting from inserting this into (2) are either elementary or can be evaluated by using (5).
 After some algebraic manipulation, we  obtain
 $$W(a,b)=4\pi^2\ln\frac{d}{2}$$
 $$+8\pi[Ti_2(a/d)+Ti_2(b/d)+Ti_2(c/d)]\eqno(6)$$
 where $d=\sqrt{ab+bc+ac}$, which is equivalent to Chen and Wu's formula.
 
 In conclusion, we list a few values of the anisotropic triangular lattice Green function that can be obtained from (6) by differentiation. Here, $\Delta(a,b,c)=a+b+c-a\cos\; x-b\cos\; y-c\cos(x+y)$, $d=\sqrt{ab+bc+ca}$.
 
 $$\int_0^{2\pi}dx\int_0^{2\pi}dy\frac{\Delta(1,0,0)}{\Delta(a,b,c)}=$$
 $$\frac{4\pi}{d^2}(b+c)[\frac{\pi}{2}+\frac{a(b+c)+2bc}{a(b+c)}\tan^{-1}(a/d)-\tan^{-1}(\frac{(b+c)d}{bc+d^2})]$$
 $$\int_0^{2\pi}dx\int_0^{2\pi}dy\frac{\Delta(0,0,1)}{\Delta(a,b,c)}=$$
 $$\frac{4\pi}{d^2}(a+b)[\frac{\pi}{2}+\frac{(a+b)c+2ab}{(a+b)c}\tan^{-1}(c/d)-\tan^{-1}(\frac{(a+b)d}{ab+d^2})]$$
 $$\int_0^{2\pi}dx\int_0^{2\pi}dy\frac{\Delta(1,1,1)}{\Delta(a,b,c)}=\frac{4\pi^2}{d^2}(a+b+c)$$
 $$+\frac{8\pi}{d^2}[\frac{bc-a^2}{a}\tan^{-1}(a/d)+\frac{ac-b^2}{b}\tan^{-1}(b/d)+\frac{ab-c^2}{c}\tan^{-1}(c/d)].$$
 $$\int_0^{2\pi}dx\int_0^{2\pi}dy\frac{\Delta(-1,1,0)}{\Delta(a,b,c)}=\frac{2\pi^2}{d^2}(a-b)$$
 $$+\frac{4\pi}{d^2}[(b-a)\tan^{-1}(c/d)+(a+b+\frac{2c(a+b)}{b})\tan^{-1}(b/d)-(a+b+2c+\frac{2c(a+b)}{a})tan^{-1}(a/d)]$$
 \vskip .2in
 
 {\bf Acknowledgement}
This work was supported in part by the Spanish MEC (BFM2002-03773 and MLG grant SAB2003-0117) and Junta de Castilla y Le¥on (VA085/02). MLG thanks the Universidad de Valladolid for hospitality and the NSF (USA) for partial support (DMR-0121146).
 
 \newpage
 \noindent {\bf References}
 
 \noindent
 [1]  Wu FY  1977 J.Phys.{\bf{A10} L113 .
 
 \noindent
 [2]  Glasser ML and Wu FY 2004 Ramanujan Journal (to appear) arXiv: cond-matter/ 0309198
 
 \noindent
 [3]  Houtappel RMF 1950 Physica {\bf 16} 425
 
 \noindent
 [4] Chen LC and Wu FY 2004 arXiv:cond-matter/0501228
 
 \noindent[5] L. Lewin, {\it Dilogarithms and Associated Functions}[1958, MacDonald \& Co, Publishers, London] p.268.

\end{document}